\documentclass[12pt,letter]{article}

\usepackage{amsmath}
\usepackage{amsfonts}
\usepackage{amsxtra}
\usepackage{amssymb}

\usepackage{graphicx,epsfig}

\newcounter{apps}

\newcounter{prs}[section]

\newcounter{cors}

\newcounter{figs}

\newcommand{\be}{\begin{equation}}
\newcommand{\ee}{\end{equation}}
\newcommand{\bea}{\begin{eqnarray*}}
\newcommand{\eea}{\end{eqnarray*}}
\newcommand{\beaa}{\begin{eqnarray}}
\newcommand{\eeaa}{\end{eqnarray}}

\newcommand{\ba}{\begin{array}}
\newcommand{\ea}{\end{array}}

\newcommand{\lb}{\label}

\newcommand{\ra}{\rightarrow}

\newcommand{\wt}{\widetilde}
\newcommand{\td}{\tilde}

\newcommand{\al}{\alpha}

\newcommand{\ld}{\lambda}
\newcommand{\Ld}{\Lambda}

\newcommand{\OO}{{\mathcal{O}}}

\newcommand{\C}{{\mathbb{C}}}
\newcommand{\NN}{{\mathbb{N}}}
\newcommand{\PP}{{\mathbb{P}}}

\begin{document}

\begin{titlepage}
\begin{center}

{\hbox to\hsize{
\hfill ITEP-TH-18/07 }}

\vspace{3cm}

{\Large \bf 
Del Pezzo singularities 
and SUSY breaking
}\\[3cm]

{\large Dmitry Malyshev${}^{a,}\footnote{On leave from ITEP Russia, Moscow, B. Cheremushkinskaya, 25}$}\\[8mm]

${}^a$ {\it Department of Physics, Princeton University, Princeton,
NJ 08544}\\[2mm]
\vspace*{2.2cm}

\end{center}

An analytic construction of compact Calabi-Yau manifolds
with del Pezzo singularities is found.
We present complete intersection CY manifolds for all del Pezzo
singularities and study the complex deformations of these
singularities.
An example of the quintic CY manifold with del Pezzo 6 singularity
and some number of conifold singularities is studied in details.
The possibilities for the 'geometric' and ISS mechanisms of
dynamical SUSY breaking are discussed.
As an example,
we construct the ISS vacuum for the del Pezzo 6 singularity.

\end{titlepage}

\newpage
\tableofcontents

\bigskip
\noindent
{\bf Appendix. A list of compact CY with del Pezzo singularities
\hspace{30mm} 27 }

\section{Motivation}

Recently there has been a substantial progress in Model building
involving the D-branes at the singularities of non compact
Calabi-Yau manifolds.
On the one hand, the singularities provide enough
flexibility to find
phenomenologically acceptable extensions of the Standard Model
\cite{Verlinde:2005jr}\cite{Wijnholt:2007vn}
and solve some problems such as finding meta-stable susy breaking
vacua \cite{Argurio:2007qk}\cite{Diaconescu:2005pc}.
On the other hand, the presence of the singularity eliminates
certain massless moduli, such as the adjoint fields on the branes
wrapping rigid cycles  \cite{Verlinde:2005jr}\cite{Wijnholt:2002qz}.

The main purpose of this paper is to study the
del Pezzo and conifold singularities on compact CY manifolds that
may be useful for the compactifications of dynamical
SUSY breaking mechanisms.
The stringy reallizations of metastable SUSY breaking vacua have
been known for some time \cite{Kachru:2002gs}\cite{Kachru:2003aw}.
We will focus on the two recent approaches to the
dynamical SUSY breaking:
on the 'geometrical' approach of
\cite{Aganagic:2006ex}\cite{Douglas:2007tu}
and on the ISS construction \cite{ISS}.
One of the main goals will be to study the topological conditions
for the compactification of the above constructions.

An important topological property of 'geometrical' mechanism is the
presence of several homologous rigid two-cycles.
This is not difficult to achieve in the case of conifold
singularities.
For example, in the geometric transitions on compact CY manifolds
\cite{Candelas:1989ug}\cite{Greene:1995hu}, several conifolds may be
resolved by a single Kahler modulus, i.e. the two-cycles at the tip of
these conifolds are homologous to each other.
However this is not always true for the del Pezzo singularities, i.e.
the two-cycles in the resolution of del Pezzo singularity may have no
homologous rigid two-cycles on the compact CY.
In the paper we explicitly construct a compact CY manifold with del
Pezzo 6 singularity and a number of conifolds such that some two-cycles
on the del Pezzo are homologous to the two-cycles of the conifolds.
This construction opens up the road for the generalization of
geometrical SUSY breaking in the case of del Pezzo singularities,
where one may hope to use the richness of deformations of these
singularity for phenomenological applications.

A more direct way towards phenomenology is provided by the
ISS mechanism.
We find an example of an ISS vacuum for the del Pezzo $6$
singularity.
A nice feature of the del Pezzo singularities is that they are
isolated.
Thus the fractional branes, that one typically introduces in these
models, are naturally stabilized against moving away from the
singularity.
But, for example, in the models involving quotients of conifolds
\cite{Argurio:2007qk}\cite{Argurio:2006ny},
the singularities are not isolated and one needs to pay special attention
to stabilize the fractional branes against moving along the singular
curves.

Apart from the application to SUSY breaking, the construction of
compact CY manifolds with del Pezzo singularities may be useful for
the study of deformations of these singularities.
In particular we will be interested in the D-brane interpretation of
deformations.

In general, a singularity can be smoothed out in two different ways,
it can be either deformed or resolved (blown up).
The former corresponds to the deformations of the complex
structure,
described by the elements of $H^{2,1}$;
the latter corresponds to K\"ahler deformations given by the
elements of $H^{1,1}$ \cite{Candelas:1990pi}\cite{GSW}\cite{Hubsch}.
In terms of the cycles, the resolution
corresponds to blowing up some
two-cycles (four-cycles)
while the complex deformations correspond to the deformations of
the three-cycles.
For example,
the conifold can be either deformed by placing an $S^3$ at the
tip of the conifold or resolved by placing an $S^2$
\cite{Candelas:1989js}.
The process where some
three-cycles shrink to form a singularity and after that the
singularity is blown up is called the geometric transition
\cite{Candelas:1989ug}\cite{Greene:1995hu}.
For the conifold, the geometric transition has a nice interpretation
in terms of the branes.
The deformation of the conifold is induced by wrapping the
D5-branes around the vanishing $S^2$ at the tip
\cite{Klebanov:2000hb}.
The resolution of the conifold corresponds
to giving a vev to a baryonic operator,
that can be interpreted in terms of the D3-branes wrapping
the vanishing $S^3$ at the tip of the conifold
\cite{Klebanov:1999tb}.

The example of the conifold encourages to conjecture that any
geometric transition can be interpreted in terms of the branes.
The non anomalous (fractional) branes
produce the fluxes that deform the three-cycles.
The massless/tensionless branes correspond to baryonic operators
whose vevs are interpreted as the blowup modes.

However, there are a few puzzles with the above interpretation. In
some cases there are less deformations than non anomalous fractional
branes, in the other cases there are deformations but no fractional
branes. The quiver gauge theory on the del Pezzo 1 singularity has a
non anomalous fractional brane, moreover it has a cascading behavior
\cite{Herzog:2004tr} similar to the conifold cascade. But it is
known that there are no complex deformations of the cone over $dP_1$
\cite{Altmann}\cite{Gross}\cite{Berenstein:2005xa}
\cite{Franco:2005zu}\cite{Bertolini:2005di}.
The relevant observation \cite{Cachazo:2001sg} is that there are no
geometric transitions for the cone over $dP_1$. From the point of
view of gauge theory, there is a runaway behavior at the bottom of
the cascade and no finite vacuum \cite{Intriligator:2005aw}.


On the other side of the puzzle, there are more complex deformations of
higher del Pezzo singularities, than there are possible
fractional branes.
It is known that the cone over del Pezzo $n$ surface has $c^\vee(E_n)-1$
complex deformations \cite{Cachazo:2001sg},
where $c^\vee(E_n)$ is the dual Coxeter number
of the corresponding Lie group.
For instance, the cone over $dP_8$ has $29$ deformations.
But there
are only $8$ non anomalous combinations of fractional branes
\cite{Verlinde:2005jr}.

We believe that these puzzles can be managed more effectively if
there were more examples of compact CY manifolds with local del Pezzo
singularities.
The advantage of working with compact manifolds is that they have
finite number of deformations and well defined cohomology (there are
no non compact cycles).

The organization of the paper is as follows.
In section 2 we construct an example of quintic CY manifold that has
both the del Pezzo and conifold singularities.
The compactness of CY manifold puts additional constrains on the
possible configurations of branes and fluxes \cite{Giddings:2001yu}.
We would like to point out that the presence of conifolds may be
necessary if we want to put fractional branes at a del Pezzo
singularity.
In our example, if the del Pezzo singularity is the only singularity
on the quintic, then all non anomalous two-cycles on del Pezzo
(i.e. the ones that don't intersect the canonical class)
turn out to be trivial within the CY manifold.
In the absence of orientifold planes we cannot put fractional branes
on such 'cycles', because on a compact manifold the RR flux from
these branes has 'nowhere to go'.
But if there are some other singularities, such as conifolds,
then it is possible that some non anomalous two-cycles on del Pezzo
are homologous to the vanishing cycles on the conifolds
(this will be the case in our example).
Then we can put
some number of D5-branes on the two-cycles of del Pezzo
and some number of anti D5-branes on the two-cycles of
the conifolds.
Such configuration of branes and anti-branes is a first step in the
geometrical SUSY breaking
\cite{Aganagic:2006ex}\cite{Heckman:2007wk}.
Also the possibility to introduce the fractional branes will be
crucial for the D-brane realizations of ISS construction.

In section 3 we discuss the compactification of the geometrical SUSY
breaking and the ISS model
and find an ISS SUSY breaking vacuum in a quiver gauge
theory for the $dP_6$ singularity.

In section 4 we formulate the general construction of compact CY
manifolds with del Pezzo singularities
and discuss the complex deformations of these singularities.
We observe that the number of deformations depends
on the global properties of the two-cycles on del Pezzo that don't
intersect the canonical class and have self-intersection (-2).
Suppose, all such cycles are trivial within the CY, then the singularity
has the maximal number of deformations.
This will be the case for our embeddings of
del Pezzo 5,6,7, and 8 singularities and
for the cone over $\PP^1\times\PP^1$.
In the case of $dP_0=\PP^2$ and $dP_1$ singularities we don't expect to
find any deformations.
In the case of del Pezzo 2,3, and 4, our embedding leaves some of the
(-2) two-cycles non trivial within the CY,
accordingly we find less complex deformations.
This result can be expected, since it is known that the del Pezzo
singularities for $n\leq 4$ in general cannot be represented as
complete intersections \cite{Gross}\cite{Reid-dP}.
In our case the del Pezzo singularities are complete intersections but
they are not generic.
Specific equations for embedding of del Pezzo singularities and
their deformations are provided in the appendix.

\nopagebreak
Section 5 contains discussion and conclusions.

\section{Del Pezzo 6 and conifold singularities on the quintic CY
\lb{seqdelp}}

The CY manifolds can have two types of primitive isolated
singularities: conifold singularities and
del Pezzo singularities \cite{Gross}\cite{Wilson}.
Correspondingly, we will have two types of geometric transitions
\begin{enumerate}
\item
Type I, or conifold transitions: several $\PP^1$'s shrink to form conifold
singularities and then these singularities are deformed.

\item
Type II, or del Pezzo transition: a del Pezzo shrinks to a
point and the corresponding singularity is deformed.
\end{enumerate}

 \begin{figure}[ht]
\begin{center}
            \scalebox{1}{
               \includegraphics[width=35pc]{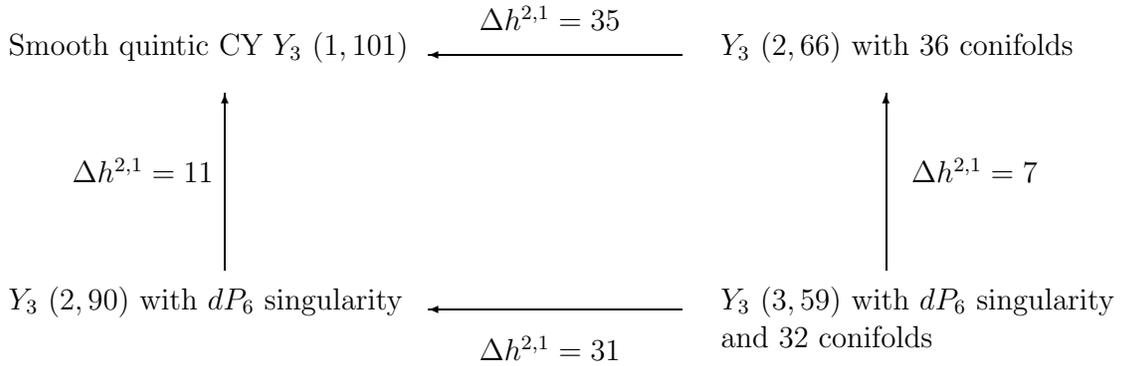}
               }

\end{center}
\vspace{-.5cm}
\caption{\it Possible geometric transitions of quintic CY.
The numbers in parentheses denote the dimensions
$(h^{1,1},\;h^{2,1})$.
}

\label{diagram}
 \end{figure}

In order to illustrate the geometric transitions we will study
a particular example of transitions on the quintic CY.
The example is summarized in the diagram in figure 1.
The type I transitions are horizontal, the type II transitions are
vertical.
It is known \cite{Cachazo:2001sg}
that the maximal number of deformations of a cone over
$dP_6$ is $c^\vee(E_6)-1=11$, where $c^\vee(E_6)=12$ is the dual
Coxeter number of $E_6$.
Going along the left vertical arrow we
recover all complex deformations of the cone over $dP_6$.
In this case all the two-cycles that don't intersect the canonical
class on $dP_6$ are trivial within the CY.

For the CY with both del Pezzo and conifold singularities,
the deformation of the del Pezzo singularity has only 7 parameters
(right vertical arrow).
The del Pezzo surface is not generic in this case.
It has a two-cycle that is non trivial within the full CY and doesn't
intersect the canonical class inside del Pezzo.
As a general rule the existence of non trivial two-cycles reduces
the number of possible complex deformations.

The horizontal arrows represent the conifold transitions.
In our example we have $36$ conifold singularities on the
quintic CY.
These singularities have $35$ complex deformations.
In the presence of $dP_6$ singularity there will be only $32$
conifolds that have, respectively, $31$ complex deformations.%
\footnote{
It may seem puzzling that we need exactly 36 or 32 conifolds.
One can easily find the examples of quintic CY with fewer conifold
singularities.
But it's impossible to blow up these singularities unless we have a
specific number of them at specific locations.
In example considered in \cite{Candelas:1989ug}\cite{Greene:1995hu},
the quintic CY has 16 conifolds placed at a $\PP^2$ inside the CY.
}

In general, the del Pezzo singularity and the conifold singularities
are away from each other but they still affect the number of complex
deformations, i.e. the presence of conifolds reduces the number of
deformations of del Pezzo singularity and vice versa.
The diagram in figure \ref{diagram} is commutative and
the total number of complex deformations of the CY with the del
Pezzo singularity and $32$ conifold singularities is $42$.
But the interpretation of these
deformations changes whether we first deform the del Pezzo
singularity or we first deform the conifold singularities.

Before we go to the calculations let us clarify what we mean by the
deformations of the del Pezzo singularity.
We will distinguish three kinds of deformations.
The deformations of the shape of the cone, the deformations of the
blown up del Pezzo with fixed canonical class and deformations that
smooth out the singularity.

The first kind of deformations corresponds to the general deformations
of del Pezzo surface at the base of the cone.
Recall that the $dP_n$ surface for $n>4$ has $2n-8$ deformations that
parameterize the superpotential of the corresponding quiver gauge
theory \cite{Wijnholt:2002qz}.

The second kind of deformations is obtained by
blowing up the singularity and fixing the
canonical class on the del Pezzo.
In this case the deformations of del Pezzo $n$ surface
can be described as
the deformations of $E_n$ singularity on the del Pezzo
\cite{Friedman:1997yq}.
The deformations of this singularity have $n$ parameters corresponding
to the $n$ two-cycles that don't intersect the canonical class.
Note, that the intersection matrix of these two-cycles is (minus)
the Cartan matrix of $E_n$.
The $E_n$ singularity on the del Pezzo is an example of du Val
surface singularity \cite{Reid-Young} (also known as an ADE
singularity or a Kleinian singularity).
A three dimensional singularity that has a du Val singularity
in a hyperplane section is called compound du Val (cDV)
\cite{Wilson}\cite{Reid-Young}.
The conifold is an example of cDV singularity since it has the $A_1$
singularity in a hyperplane section.
The generalized conifolds \cite{Gubser:1998ia}\cite{Corrado:2004bz}
also have an ADE
singularity in a hyperplane section, i.e. from the 3-dimensional point of view
they correspond to some cDV singularities.
In terms of the large $N$ gauge/string duality the
deformation of the $E_n$ generalized conifold singularity
corresponds to putting some
combination of fractional branes on the zero size two-cycles at the
singularity.
Hence the deformtion of cDV singularity that restricts to
$E_n$ singularity on the del Pezzo can be
considered as a generalized type I transition.

We will be mainly interested in the
the third type of deformations that correspond to smoothing of
del Pezzo singularities.
These deformations make the canonical class of del Pezzo surface
trivial within the CY.
If we put some number of non anomalous fractional D-branes at
the singularity, then the corresponding geometric transition smooths
the singularity \cite{Cachazo:2001sg}.
But not all the deformations can be described in this way.

In order to get some intuition about possible interpretations of
these deformations we will consider the del Pezzo 6 singularity.
It is known that the $dP_6$ singularity has $11$ complex deformations
\cite{Altmann}\cite{Kapustka}
but there are only $6$ non anomalous fractional branes in the
corresponding quiver gauge theory and there are only $6$ two-cycles
that don't intersect the canonical class \cite{Cachazo:2001sg}.
It will prove helpful to
start with a quintic CY that has $36$ conifold singularities.
The del Pezzo $6$ singularity can be obtained by merging four
conifolds at one point.
There are $7$ deformations of del Pezzo 6 singularity that separate
these four conifolds (right vertical arrow).
The remaining $4$ deformations of $dP_6$ cone correspond to
$4$ deformations of the four "hidden" conifolds at the singularity.
Note, that the total number of deformations is 11 (left vertical
arrow).

\subsection{Quintic CY}

The description of the quintic CY is well known \cite{Hubsch}.
Here we repeat it in order to recall the methods \cite{Hubsch}
of finding the
topology and deformations that we use later in more difficult
situations.

The quintic CY
manifold $Y_3$ is given by a degree five equation in $\PP^4$
\be\lb{quint}
Q_5(z_i)=0
\ee
where $(z_0,z_1,z_2,z_3,z_4)\in\PP^4$.
The total Chern class of this manifold is
\be
c(Y_3)=\frac{(1+H)^5}{1+5H}=1+10H^2-40H^3
\ee
the first Chern class $c_1(Y_3)=0$.

Let us calculate the number of complex deformations.
The complex structures are parameterized by the coefficients in
(\ref{quint}) up to the change of coordinates in $\PP^4$.
The number of coefficients in a homogeneous polynomial
of degree $n$ in $k$ variables is
\be
(^{n+k-1}_n)=\frac{(n+k-1)!}{n!(k-1)!}
\ee
In the case of the quintic in $\PP^4$ the number of coefficients is
\be
(^{9}_5)=\frac{9!}{5!4!}=126
\ee
The number of reparametrizations of $\PP^4$ is equal to
${\rm dim} Gl(5)=25$.
Thus the dimension of the space of complex deformations is $101$.

The number of complex deformations of CY threefolds is equal to
the dimension of $H^{2,1}$ cohomology group
\be
h^{2,1}=h^{1,1}-\chi/2,
\ee
where $h^{1,1}$ can be found via the Lefschetz hyperplane theorem
\cite{Hubsch}\cite{Griffiths}
\be
h^{1,1}(Y_3)=h^{1,1}(\PP^4)=1
\ee
and the Euler characteristic
is given by the integral of the highest Chern class over $Y_3$
\be
\chi=\int_{Y_3}c_3=\int_{\PP^4}-40H^3\wedge 5H=-200,
\ee
here we have used that $5H$ is the Poincare dual class to $Y_3$
inside $\PP^4$.
Consequently $h^{2,1}=101$ which is consistent with the number of
complex deformations found before.

\subsection{Quintic CY with $dP_6$ singularity}

Suppose that the quintic polynomial is not generic but has a degree
three zero at the point $(w_0,w_1,w_2,w_3,w_4)=(0,0,0,0,1)$
\be\lb{quidep6}
P_3(w_0,\ldots,w_3)w_4^2
+P_4(w_0,\ldots,w_3)w_4
+P_5(w_0,\ldots,w_3)=0
\ee
where $P_n$'s denote degree $n$ polynomials.
The shape of the singularity is determined by
$P_3(w_0,\ldots,w_3)$,
(we will see that this polynomial defines the del Pezzo
at the tip of the cone).
The deformations that smooth out the singularity
correspond to adding less singular terms to
(\ref{quidep6}), i.e. the terms that have bigger powers of $w_4$.


The resolution of the singularity in (\ref{quidep6}) can be obtained
by blowing up the point $(0,0,0,0,1)\in\PP^4$.
Away from the blowup we can use the
following coordinates on ${\PP}^4$
\be\lb{parpin0}
(w_0,\ldots,w_3,w_4)=(tz_0,\ldots,tz_{3},s)
\ee
where $(s,t)\in \PP^1$ and $(z_0,\ldots,z_3)\in\PP^3$.
The blowup of the point at $t=0$ corresponds to inserting the $\PP^3$
instead of this point.
Hence the points on the blown up ${\PP}^4$ can be
parameterized globally by
$(z_0,\ldots,z_3)\in\PP^3$ and $(s,t)\in\PP^1$.
The projective invariance $(s,t)\sim (\ld s,\ld t)$ corresponds to
the projective invariance in the original $\PP^4$.
In order to compensate for the projective invariance of $\PP^3$ we
need to assume that
locally the coordinates on $\PP^1$ belong to the following
line bundles over $\PP^3$,
$s\in\OO$ and $t\in\OO(-H)$.
Thus the blowup of $\PP^4$ at a point is a
$\PP^1$ bundle over $\PP^3$ obtained by projectivization of the
direct sum of $\OO_{\PP^3}$ and $\OO_{\PP^3}(-H)$ bundles,
$\wt{\PP}^4=P(\OO_{\PP^3}\oplus\OO_{\PP^3}(-H))$
(for more details on projective bundles see,
e.g. \cite{Morrison:1996na}\cite{Andreas:1998zf}).
In working with projective bundles, we will use the technics similar to
\cite{Andreas:1998zf}.

Using parametrization (\ref{parpin0}), we can write the equation on the
blown up $\PP^4$ as
\be\lb{eqdep6}
P_3(z_0,\ldots,z_3)s^2
+P_4(z_0,\ldots,z_3)st
+P_5(z_0,\ldots,z_3)t^2=0.
\ee
This equation is homogeneous of degree two in the
coordinates on $\PP^1$ and degree three in the $z_i$'s.
Note, that $t\in\OO(-H)$, i.e. it has degree $(-1)$ in the $z_i$'s, and
$s\in\OO$ has degree zero.

Let us prove that the manifold defined by (\ref{eqdep6}) has
vanishing first Chern class, i.e. it is a CY manifold.
Let $H$ be the hyperplane class in $\PP^3$ and $G$ be the hyperplane
class on the $\PP^1$ fibers.
Let $M=P(\OO_{\PP^3}\oplus\OO_{\PP^3}(-H))$ denote the
$\PP^1$ bundle over $\PP^3$.
The total Chern class of $M$ is
\be
c(M)=(1+H)^4(1+G)(1+G-H)
\ee
where $(1+H)^4$ is the total Chern class of $\PP^3$,
$(1+G)$ corresponds to $s\in\OO_{\PP^3}$ and
$(1+G-H)$ corresponds to $t\in\OO_{\PP^3}(-H)$.
Note, that $G(G-H)=0$ on this $\PP^1$ bundle and,
as usual, $H^4=0$ on the $\PP^3$.

Let $Y_3$ denote the surface embedded in $M$ by (\ref{eqdep6}).
Since the equation has degree $3$ in $z_i$ and degree two in
$(s,t)$,
the class Poincare dual to $Y_3\subset M$ is $3H+2G$ and the total
Chern class is
\be\lb{qcdp6}
c(Y_3)=\frac{(1+H)^4(1+G)(1+G-H)}{1+3H+2G}.
\ee
Expanding $c(Y_3)$, it is easy to check that
$c_1(Y_3)=0$.

The intersection of $Y_3$ with the blown up $\PP^3$ at $t=0$ is
given by the degree three equation $P_3(z_0,\ldots,z_3)=0$ in $\PP^3$.
The surface $B$ defined by this equation is the del Pezzo 6
surface \cite{Hubsch}\cite{Griffiths}.
The total Chern class and the Euler character of $B$
\beaa
\lb{ccdp6}
c(B)=\frac{(1+H)^4}{1+3H}=1+H+3H^2;\\
\chi(B)=\int_B c_2(B)=\int_{\PP^3}3H^2\wedge 3H=9.
\eeaa
In the calculation of $\chi(B)$ we have used that $3H$ is the
Poincare dual class to $B$ inside $\PP^3$.

It is known that the normal bundle to contractable del Pezzo in a
CY manifold is the canonical bundle on del Pezzo \cite{Buican:2006sn}.
Let us check this statement in our example.
The canonical class is minus the first Chern class that can be found
from (\ref{ccdp6})%
\footnote{
Slightly abusing the notations, we denote by $H$ both the class of
$\PP^3$ and the restriction of this class to $B\in\PP^3$.
}
\be\lb{cancl}
K(B)=-H.
\ee
The coordinate $t$ describes the normal direction to $B$ inside $Y_3$.
Since $t\in\OO_{\PP^3}(-H)$, restricting to $B$ we find that $t$
belongs to the canonical bundle over $B$.
Hence locally, near $t=0$, the CY threefold $Y_3$ has
the structure of the CY cone over the del Pezzo 6 surface.


The smoothing of the singularity corresponds to adding less singular
terms in (\ref{quidep6}).
These terms have $15$ parameters, but also we get back $4$
reparametrizations (now we can add $w_4$ to the other coordinates).
Hence smoothing of the singularity corresponds to $11$ complex
structure deformations that
is the maximal expected number of deformations of $dP_6$
singularity.

In view of applications in section 4 let us describe the
geometric transition between the CY with the resolved
 $dP_6$ singularity and a smooth quintic CY in more details.
As we have shown above,
the CY with the blown up $dP_6$ singularity can be described by the
following equation in the $\PP^1$ bundle over $\PP^3$
\be
P_3(z_0,\ldots,z_3)s^2
+P_4(z_0,\ldots,z_3)st
+P_5(z_0,\ldots,z_3)t^2=0
\ee
This equation can be rewritten as
\be\lb{quinnew}
P_3(tz_0,\ldots,tz_3)s^2
+P_4(tz_0,\ldots,tz_3)s
+P_5(tz_0,\ldots,tz_3)=0
\ee
Next we note that, being a projective bundle, $M$ is equivalent
\cite{Griffiths}\cite{Morrison:1996na} to
$P(\OO_{\PP^3}(H)\oplus\OO_{\PP^3})$, where locally $s$ and $t$ are sections of
$\OO_{\PP^3}(H)$ and $\OO_{\PP^3}$ respectively.
We further observe that $tz_i,\;i=0\ldots 3$ are also sections of
$\OO_{\PP^3}(H)$
and the equivalence $(t,s)\sim (\ld t, \ld s)$ induces the
equivalence
$(tz_0,\ldots,tz_i,s)\sim (\ld tz_0,\ldots,\ld tz_i,\ld s)$.
Consequently, if we blow down the section
$t=0$ of $M$, then $(tz_0,\ldots,tz_i,s)\in\PP^4$.
Now we define $(w_0,\ldots,w_3,w_4)=(tz_0,\ldots,tz_3,s)$ and rewrite
(\ref{quinnew}) as
\be\lb{quinnew2}
P_3(w_0,\ldots,w_3)w_4^2
+P_4(w_0,\ldots,w_3)w_4
+P_5(w_0,\ldots,w_3)=0
\ee
Not surprisingly, we get back equation (\ref{quidep6}).

Above we have found that there are $11$ complex deformations of the
$dP_6$ singularity embedded in the quintic CY manifold.
In the view of further applications let us rederive the number of
complex deformations by calculating the dimension of $H^{2,1}$.

Expanding (\ref{qcdp6}), we get the third Chern class
\be
c_3(Y_3)=-2G^3-13HG^2-17H^2G-8H^3.
\ee
The Poincare dual class to $Y_3\in M$ is $3H+2G$ and
\be
\chi(Y_3)=\int_{Y_3}c_3(Y_3)=\int_{M}c_3(Y_3)\wedge(3H+2G).
\ee
In calculating this integral one needs to take into account that
$G(G-H)=0$ on $M$.
Finally we get
\be
\chi(Y_3)=-176
\ee
and
\be
h^{2,1}=h^{1,1}-\chi/2=90.
\ee
The number of complex deformations of the del Pezzo
singularity is $101-90=11$, which is consistent with the number
found above.

\subsection{Quintic CY with $36$ conifold
singularities}


In this subsection we use the methods of geometric transitions
\cite{Candelas:1989ug}\cite{Greene:1995hu}\cite{Hubsch}
to find the quintic CY with
conifold singularities, i.e. we describe the upper horizontal
arrow in figure \ref{diagram}.
Consider the system of two equations in $\PP^4\times\PP^1$
\be\lb{squin}
\left\{
\ba{lll}
P_3u+R_3v=0\\
P_2u+R_2v=0
\ea
\right.
\ee
where $(u,v)\in\PP^1$ and $P_n$, $R_n$ denote polynomials of degree
$n$ in $\PP^4$.

Suppose that at least one of the polynomials $P_3, R_3, P_2$ and $R_2$
is non zero, then we can solve for $u,v$ and substitute in the
second equation, where we get
\be\lb{nquin}
P_3R_2-R_3P_2=0
\ee
a non generic quintic in $\PP^4$.
The points where $P_3=R_3=P_2=R_2=0$ (but otherwise generic)
have conifold singularities.
There are $3\cdot 3\cdot 2\cdot 2=36$ such points.
The system (\ref{squin}) describes the blowup of the
singularities, since every singular point is replaced by the $\PP^1$
and the resulting manifold is non singular.

Let $H$ be the hyperplane class of $\PP^4$ and $G$ by the hyperplane
class of $\PP^1$, then the total Chern class of $Y_3$ is
\be\lb{chnquin}
c=\frac{(1+H)^5(1+G)^2}{(1+3H+G)(1+2H+G)},
\ee
since $c_1=0$, $Y_3$ is a CY.

By Lefschetz hyperplane theorem
$h^{1,1}(Y_3)=h^{1,1}(\PP^4\times\PP^1)=2$, there are only two
independent Kahler deformations in $Y_3$. One of them is the overall
size of $Y_3$ and the other is the size of the blown up $\PP^1$'s.
Thus the $36$ $\PP^1$'s are not independent but homologous to each
other and represent only one class in $H_2(Y_3)$.
If we shrink the size of blown up $\PP^1$'s to zero, then we can
deform the singularities of (\ref{nquin}) to get a generic quintic
CY.
In this case the 35 three chains that where connecting the 36
$\PP^1$'s become independent three cycles.
Thus we expect the general quintic CY to have 35 more complex
deformations than the quintic with $36$ conifold singularities.

Calculating the Euler character similarly to the
previous subsections, we find
\be
h^{2,1}=66.
\ee
Recall that the smooth quintic has $101$ complex deformations.
Thus the quintic with $36$ conifold singularities has
$101-66=35$ less complex deformations than the generic one.

\subsection{Quintic CY with $dP_6$ singularity
and $32$ conifold singularities}

The equation for the quintic CY manifold with the blown up $dP_6$
singularity was found in (\ref{eqdep6}).
Here we reproduce it for convenience
\be\lb{dp6def}
P_3(z_i)s^2+P_4(z_i)st+P_5(z_i)t^2=0
\ee
This equation describes an embedding of the CY manifold in the
$\PP^1$ bundle $M=P(\OO_{\PP^3}\oplus\OO_{\PP^3}(-H))$.
As before,
$(z_0,\ldots,z_3)\in\PP^3$ and $(s,t)$ are the coordinates on
the $\PP^1$ fibers over $\PP^3$.

In order to have more Kahler deformations we need to embed
(\ref{dp6def}) in a space with more independent two-cycles.
For example, we can consider a system of two equations in the
product ($\PP^1$ bundle over $\PP^3$) $\times$ $\PP^1$
\be\lb{dp6syst}
\left\{
\ba{lll}
(P_1s+P_2t)u+(Q_1s+Q_2t)v=0\\
(R_2s+R_3t)u+(S_2s+S_3t)v=0
\ea
\right.
\ee
where $(u,v)$ are the coordinates on the additional $\PP^1$.
Let $G$, $H$, and $K$
be the hyperplane classes on the $\PP^1$ fibers, on the
$\PP^3$, and on the additional $\PP^1$ respectively.
Then the total Chern class of $Y_3$ is
\be
c=\frac{(1+H)^4(1+G)(1+G-H)(1+K)^2}{(1+H+G+K)(1+2H+G+K)}
\ee
and it's easy to see that the first Chern class is zero.

For generic points on the $\PP^1$ bundle over $\PP^3$
at least one of the functions in front of $u$ or $v$ is non zero.
Thus we can find a point $(u,v)$ and substitute it in the second
equation, which becomes a non generic equation similar to
(\ref{dp6def})
\be\lb{dp6defnew}
(P_1S_2-Q_1R_2)s^2+(P_1S_3+P_2S_2-Q_1R_3-Q_2R_2)st
+(P_2S_3-Q_2R_3)t^2=0.
\ee

The CY manifold defined in (\ref{dp6syst}) has the following
characteristics
\bea
\chi=\int_{Y_3}c_3=-112;\\
h^{1,1}=3;\\
h^{2,1}=h^{1,1}-\chi/2=59.
\eea
Recall that the number of complex deformations on the quintic with
the del Pezzo 6 singularity is $90$.
Since we lose $31$ complex deformations we expect that the
corresponding three-cycles become the three chains that connect $32$
$\PP^1$'s at the blowups of the singularities in (\ref{dp6defnew}).
These singularities occur when all four equations in
(\ref{dp6syst}) vanish
\bea
R_2s+R_3t=0\\
S_2s+S_3t=0\\
P_1s+P_2t=0\\
Q_1s+Q_2t=0
\eea
The number of solutions of these equations
equals the number of intersections of
the corresponding classes
$\int_M(2H+G)^2(H+G)=32$, where $M$ is the $\PP^1$ bundle over
$\PP^3$ and $G(G-H)=0$.

The right vertical arrow corresponds to smoothing of del Pezzo
singularity in the presence of conifold singularities.
Before the transition the CY has $h^{2,1}=59$ deformations and after
the transition it has $h^{2,1}=66$ deformations.
Hence the number of complex deformations of $dP_6$ singularity is
$66-59=7$ which is less than $c^\vee(E_6)-1=11$.
This is related to the fact that the del Pezzo at the tip of the
cone is not generic.
The equation of the del Pezzo can be found by restricting (\ref{dp6syst}) to
$t=0$, $s=1$ section
\be\lb{newdp6}
\left\{
\ba{lll}
P_1u+Q_1v=0\\
R_2u+S_2v=0
\ea
\right.
\ee
This del Pezzo contains a two-cycle $\al$ that
is non trivial within the full CY and doesn't intersect the
canonical class inside $dP_6$.

In the rest of this subsection we will argue that $\al$ is
homologous to four $\PP^1$'s at the tip of the conifolds.
The heuristic argument is the following.
The formation of $dP_6$ singularity on the CY manifold with $36$
conifolds reduces the number of conifolds to $32$.
Let us show that the deformation of the del Pezzo singularity that
preserves the conifold singularities corresponds to separating $4$
conifolds hidden in the del Pezzo singularity.
The CY that has a $dP_6$ singularity and $32$
resolved conifolds can be found from (\ref{dp6syst}) by
the following coordinate redefinition
$(w_0,\ldots,w_3,w_4)=(tz_0,\ldots,tz_3,s)$ (compare to the
discussion after equation (\ref{quinnew}))
\be\lb{dp6syst2}
\left\{
\ba{lll}
(P_1w_4+P_2)u+(Q_1w_4+Q_2)v=0\\
(R_2w_4+R_3)u+(S_2w_4+S_3)v=0
\ea
\right.
\ee
If we blow down the $\PP^1$, then we get the quintic CY with
$32$ conifold singularities and a $dP_6$ singularity.
For a finite size $\PP^1$, the conifold singularities and one of the
two-cycles in the $dP_6$ are blown up.
The deformations of $dP_6$ singularity correspond to adding terms
with higher power of $w_4$.
After the deformation, the degree two zeros of $R_2$ and $S_2$
will split into four degree one zeros that correspond to the four
conifolds "hidden" in the $dP_6$ singularity.
The blown up two-cycle of $dP_6$ is homologous to the
two-cycles on the four conifolds.%
\footnote{
Formally we can prove this by calculating the corresponding Poincar\'e
dual classes.
The Poincar\'e dual of $\PP^1$ on the blown up conifold is $H^3G$ --
this is the $\PP^1$ parameterized by $(u,v)$.
The Poincar\'e dual of the canonical class on $dP_6$ is
$(G-H)(H+K)(2H+K)(-H)$, where $(G-H)$ restricts to $t=0$ section of
the $\PP^1$ bundle, $(H+K)(2H+K)$ restricts to $dP_6$ in
(\ref{newdp6}), while the restriction of $(-H)$ is the canonical
class on $dP_6$ (see Eq. (\ref{cancl})).
The class $\al$ that doesn't intersect $(-H)$ inside $dP_6$ is dual to
$(G-H)(H+K)(2H+K)(2H-3G)=4H^3G$, q.e.d.
}



\section{SUSY breaking}

In the paper we compare
two mechanisms for dynamical SUSY breaking:
the 'geometrical' approach of Aganagic et al \cite{Aganagic:2006ex}
and a more 'physical' approach of ISS \cite{ISS}.

In both approaches there is a confinement in the microscopic gauge
theory leading to  the SUSY breaking in the effective theory.
But the particular mechanisms and the effective theories are quite
different.
In the 'geometrical' approach the effective theory is a non SUSY
analog of Veneziano-Yankielowicz superpotential \cite{Veneziano}
for the gaugino bilinear field $S$.
This potential has an interpretation as the GVW superpotential
\cite{GVW} for the complex structure moduli of the CY
manifold.
The original Veneziano-Yankielowicz potential \cite{Veneziano}
is derived for the pure YM theory without any flavors.
It has a number of isolated vacua and no massless fields.
This is a nice feature for the (meta) stability of the vacuum but,
since all the fields are massive, the applications of this potential
in the low energy effective theories are limited
(see e.g. the discussion in \cite{intril94}).

In the ISS construction the number of flavors is bigger than the
number of colors $N_c<N_f<3/2 N_c$ (and probably $N_f=N_c$).
After the confinement the low energy effective theory contains
classically massless fields that get some masses only at 1 loop.
Hence this theory is a more genuine effective theory but the geometric
interpretation is harder to achieve
\cite{Argurio:2007qk}\cite{Diaconescu:2005pc}.
Moreover the geometric constructions similar to
\cite{Argurio:2007qk}\cite{Diaconescu:2005pc}
generally have D5-branes wrapping vanishing cycles.
In any compactification of these models, one has to put the O-planes
or anti D5-branes somewhere else in the geometry, i.e. the analysis of
\cite{Aganagic:2006ex}\cite{Douglas:2007tu} becomes inevitable.

In summary, it seems that the ISS construction is more useful for
immediate applications to SUSY breaking in the low energy effective
theories, whereas more global geometric analysis of
\cite{Aganagic:2006ex}\cite{Douglas:2007tu} becomes inevitable in the
compactifications.

In the previous section we constructed the compact CY with del Pezzo 6
singularity and some number of conifold singularities.
We have shown that it's possible to make some two-cycles on del Pezzo
homologous to the two-cycles on the conifolds.
This is the first step in the geometric analysis of
\cite{Aganagic:2006ex}.
In the next subsection we show how
the ISS story can be represented in the del Pezzo 6 quiver gauge
theories.

\subsection{ISS vacuum for the $dP_6$ singularity}

 \begin{figure}[ht]
\begin{center}
            \scalebox{1}{
               \includegraphics[width=20pc]{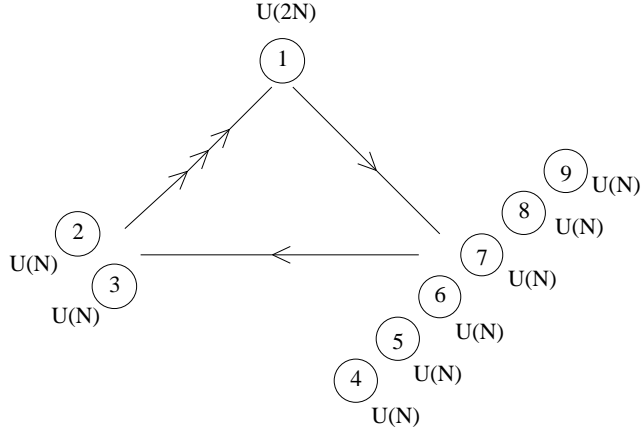}
               }
\end{center}
\vspace{-.5cm}
\caption{\it Quiver gauge theory for the cone over $dP_6$.}
\label{dP6-initial}
 \end{figure}

Consider the quiver gauge theory for the cone
over $dP_6$ represented in figure \ref{dP6-initial}.
This quiver can be found by the standard methods
\cite{Verlinde:2005jr} from the three-block exceptional
collection of sheaves \cite{Nogin}.
But, in order to prove the existence of this quiver,
it is easier to do the Seiberg
dualities on the nodes 4,5,6 and 1 and reduce it
to the known $dP_6$ quiver \cite{Wijnholt:2007vn}.

On compact CY manifolds, it is possible to have D5-branes
only in the presence of specific orientifolds or anti branes wrapping
homologous cycles somewhere else in the geometry.
In the previous section we have found a non anomalous two-cycle $\al$ on del
Pezzo $6$ that is homologous to the two-cycles of the conifolds.

\begin{figure}[ht]
\begin{center}
            \scalebox{1}{
               \includegraphics[width=20pc]{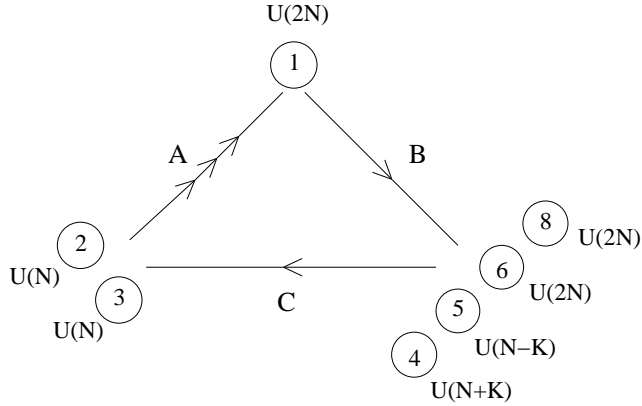}
               }
\end{center}
\vspace{-.5cm}
\caption{\it Quiver gauge theory for the cone over $dP_6$ after adding
the fractional branes.}
\label{dP6-second}
 \end{figure}

Let $A_i$ denote the two-cycle corresponding to the
D5-brane charge \cite{Verlinde:2005jr} of the bound
state of branes at the $i$-th node in figure \ref{dP6-initial}.
Note that the cycles $A_4-A_5$, $A_6-A_7$, and $A_8-A_9$ correspond to
non-anomalous $U(1)$ symmetries.
We will assume that it is possible to construct a compact CY manifold
such that these cycles
are homologous to some two-cycles
on the conifolds (or some other singularities away from del Pezzo).
Now we would like to add $K$ fractional branes to
$A_4-A_5$ and $N$ fractional branes to
$A_6-A_7$ and to $A_8-A_9$.
The corresponding quiver is depicted in figure \ref{dP6-second}.

The gauge groups at the nodes $6$ and $8$ have $N_f=N_c$.
Consider the
Seiberg duality in the strong coupling limit of these gauge groups.
The moduli space consists of the mesonic and the baryonic branches
\cite{Seiberg:1994bz}\cite{Intriligator:1995au}.
Suppose we are on the baryonic branch.
For the generic Yukawa couplings,
the two mesons $\Phi=BC$ couple linearly to the fields
$A$ and become massive together with two of the $A$ fields.

\begin{figure}[ht]
\begin{center}
            \scalebox{1}{
               \includegraphics[width=20pc]{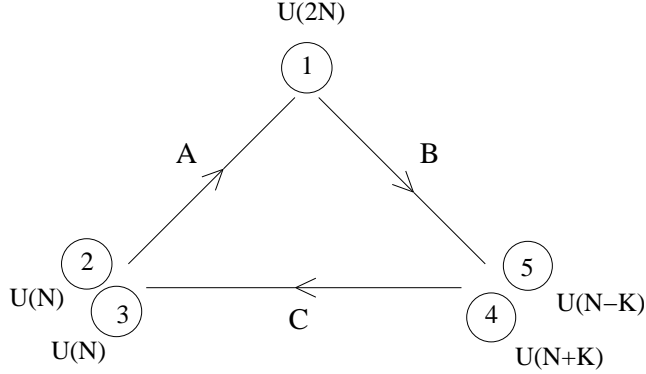}
               }
\end{center}
\vspace{-.5cm}
\caption{\it Quiver gauge theory for the cone over $dP_6$ after confinement
of nodes $6$ and $8$.}
\label{dP6-third}
 \end{figure}

An important question is whether the baryons for the gauge groups in
nodes $6$ and $8$ remain massless.
The baryons are charged under the baryonic $U(1)_B$ symmetries.
In the non compact setting these $U(1)_B$ symmetries are global
\cite{Gubser:2004qj}.
If the baryons get vevs, then the symmetries are broken spontaneously
and there are massless goldston bosons.
But for the compact CY manifold the $U(1)_B$ symmetries are gauged and the
goldstone bosons become massive \cite{Argurio:2006ny}\cite{Gubser:2004qj}
through the Higgs mechanism.
Integrating out the massive fields we get the quiver in
figure \ref{dP6-third}.

\begin{figure}[ht]
\begin{center}
            \scalebox{1}{
               \includegraphics[width=16pc]{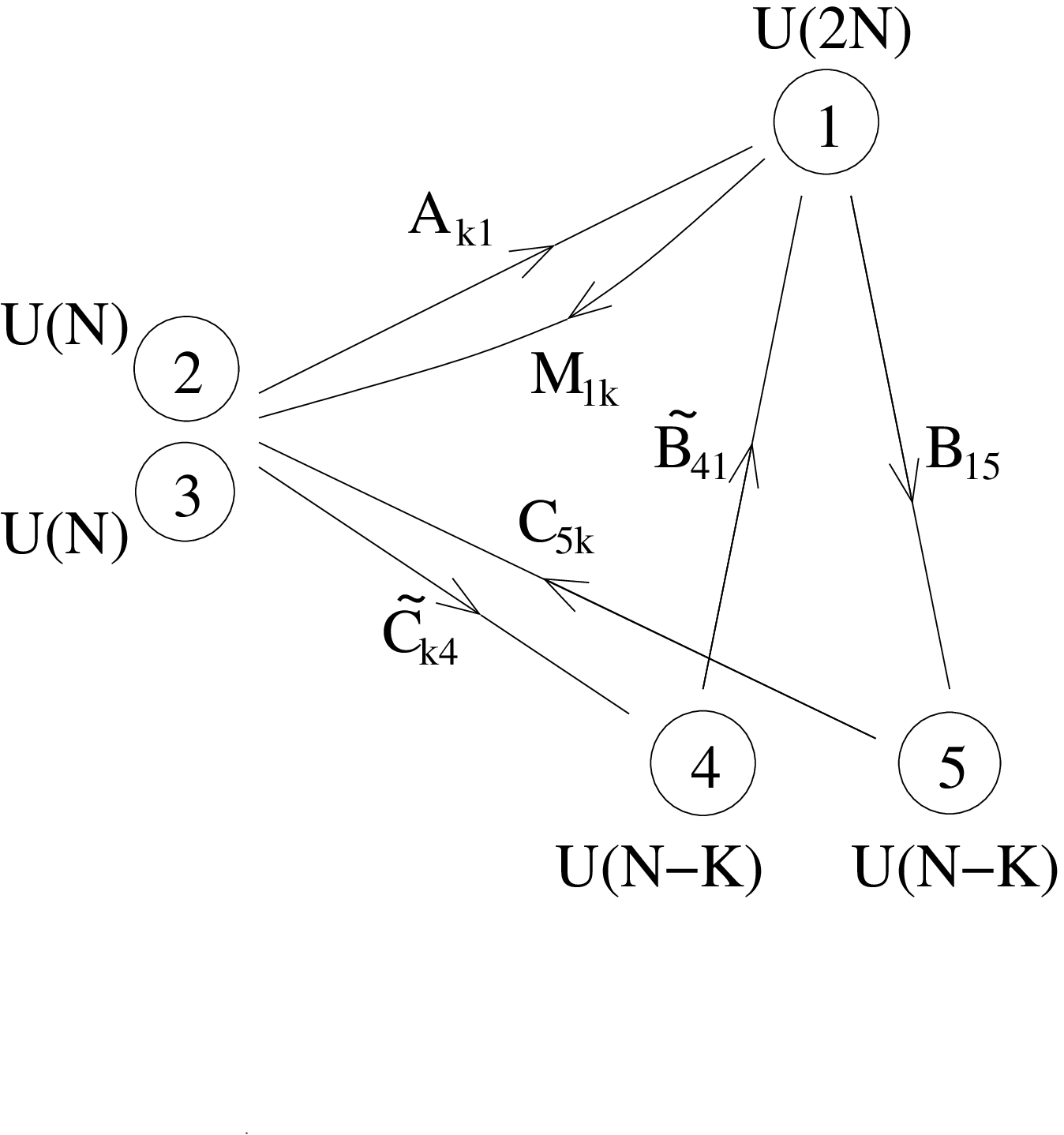}
               }
\end{center}
\vspace{-2.3cm}
\caption{\it Quiver gauge theory for the cone over $dP_6$ after
Seiberg duality on node $4$.
}
\label{dP6-fourth}
 \end{figure}

Next we assume that the strong coupling scale for the gauge group
$SU(N+K)$ at node $4$ is bigger than the scale for the
$SU(2N)$.
This assumption doesn't include a lot of tuning especially if
$K\lesssim N$.
The number of flavors for the gauge group $SU(N+K)$ is
$N_f=2N > N_c=N+K$.
Consequently, we can assume that the
mesons don't get VEVs after the confinement of $SU(N+K)$ and remain
massless.
The corresponding quiver is shown in figure \ref{dP6-fourth}.
The subscripts of the bifundamental fields denote the gauge groups
at the ends of the corresponding link.
The subscript $k=2,3$ labels the two $U(N)$ gauge groups on the left.
For example, $A_{k1}$ denotes both the field $A_{21}$ going from the
node $2$ to the node $1$ and $A_{31}$ going from $3$ to $1$.

The superpotential of the quiver gauge theory in figure
\ref{dP6-fourth} has the form
\be
\ba{lll}
W&=&{\rm Tr}(m A_{21}M_{12}+m A_{31}M_{13})\\
&+&{\rm Tr}(
\ld M_{12}\td{C}_{24}\td{B}_{41}
+\ld A_{21}B_{15}C_{52}
+\ld M_{13}\td{C}_{34}\td{B}_{41}
+\ld A_{31}B_{15}C_{53})
\ea
\ee
In order to make the notations shorter,
we don't write the subscripts of the couplings.
(The couplings are different but have the same
order of magnitude.)

If $\Ld_{1}$ for the $SU(2N)$ gauge group at node $1$ is close to
$\Ld_{4}$ for $SU(N+K)$ at node $4$ in figure \ref{dP6-third}, then
it is natural to assume that for small values of corresponding
Yukawa couplings the mass parameters $m$
satisfy $m<<\Ld_{1}$.
Now we note that the $SU(2N)$ gauge group
has $N_c=2N$ and $N_f=3N-K$, i.e. $N_c+1\leq N_f < 3/2 N_c$.
This group is a good candidate for
the microscopic gauge group in the ISS construction.
After the Seiberg duality, the magnetic gauge group has
$\td{N}_c=N-K$.
The superpotential of the dual theory is
\be
\ba{lll}
\td W&=&{\rm Tr}(m M_{22}+m M_{33})\\
&+&{\rm Tr}(\ld M_{22}\td{M}_{21}\td{A}_{12}
+\ld M_{33}\td{M}_{31}\td{A}_{13})\\
&+&{\rm Tr}(
m {M}_{42}\td{C}_{24}
+m M_{25}C_{52}
+m {M}_{43}\td{C}_{34}
+m M_{35}C_{53})\\
&+&{\rm Tr}(
  \ld M_{42} \td{M}_{21}\td{\td{B}}_{14}
+ \ld M_{25} \td{B}_{51}\td{A}_{12}
+ \ld M_{43} \td{M}_{31}\td{\td{B}}_{14}
+ \ld M_{35} \td{B}_{51}\td{A}_{13}
)
\ea
\ee
The indices of the meson fields correspond to the two gauge groups
under which they transform.
In our case this leads to unambiguous identifications,
for example, $M_{22}=A_{21}M_{12}$, $M_{33}=A_{31}M_{13}$,
$M_{42}=\td{B}_{41}M_{12}$ etc.
The mesons $M_{22}$ and $M_{33}$ are in adjoint representation of
$SU(N)_2$ and $SU(N)_3$, their F-term equations read
\be\lb{vac}
\ba{l}
m\cdot {\rm\bf 1}+\ld \td{M}_{21}\td{A}_{12}=0\\
m\cdot {\rm\bf 1}+\ld \td{M}_{31}\td{A}_{13}=0
\ea
\ee
here ${\rm\bf 1}$ is the $N\times N$ identity matrix.
The Seiberg dual gauge group at node $1$ is $SU(N-K)$,
hence the rank of
the matrices $\td{M}_{21}$ etc. is at most $N-K$
and the SUSY is broken by the rank condition of
\cite{ISS}.
Classically, there are massless excitations around the vacua in
(\ref{vac}).
In order to prove that the vacuum is metastable one has to
check that these fields acquire a positive mass at 1 loop.
Similarly to \cite{ISS} we expect this to be true, but a more
detailed study is necessary.

As a summary, in this section we have found an example of dymanical
SUSY breaking in the quiver gauge theory on del Pezzo singularity.
An interesting property of this example is that there are massless
chiral fields after the SUSY breaking.
This behavoir seems to be quite generic and we expect that similar
constructions are possible for other del Pezzo singularities.

\section{Compact CY manifolds with del Pezzo
singularities}

The {\it non compact} CY manifolds with del Pezzo singularities are known
\cite{Gross}\cite{Reid-dP}.
The $dP_n$ singularities for $5\leq n \leq 8$ and for the cone over
$\PP^1\times\PP^1$ can be represented as
complete intersections.%
\footnote{
Note, that in mathematics literature the del Pezzo surfaces are
classified by their degree $k=9-n$, where $n$ is the number of blown
up points in $\PP^2$.}
The CY cones over $\PP^2$ and $dP_n$ for $1\leq n\leq 4$ are not
complete intersections.
The {\it compact} CY manifolds for complete intersection singularities where
presented in \cite{Kapustka}.
The construction of elliptically fibred compact
CY manifolds with del Pezzo
singularities can be found e.g. in \cite{Diaconescu:2005pc}.

In our construction we use both the methods of complete intersection
CY manifolds \cite{Hubsch}
and the methods of spherical/elliptic fibrations similar to
\cite{Diaconescu:2005pc}.
Recall the construction of compact CY manifolds with local del Pezzo
singularities via elliptic fibrations \cite{Diaconescu:2005pc}.
The first step is to take a particular $\PP^1$ bundle over the del
Pezzo.
The resulting threefold $B_3$ can be viewed as a base for the
F-theory CY fourfold.
In the type IIB limit of F-theory the CY fourfold becomes a
CY threefold that has the form of a double cover of $B_3$.
This double cover of the $\PP^1$ bundle is an elliptic fibration over the
del Pezzo.

In our construction we first embed
the del Pezzo surface $B$ in a space $X$,
where $X$ is a projective space, a product of projective spaces,
or a weighted projective space \cite{Hubsch}\cite{Friedman:1997yq}.
Then we consider a particular $\PP^1$ bundle over the space $X$
(not only over the del Pezzo).
The CY threefold $Y_3$ is embedded in this $\PP^1$ bundle via a
complete intersection of a system of equations.
One of the sections of the $\PP^1$ bundle is contractible and
intersects $Y_3$ by the del Pezzo surface.
The contraction of this section corresponds to forming the del
Pezzo singularity on the CY manifold.
The description as a system of equations enables one to identify more
easily the complex deformations of the singularity than in the case of
elliptic fibrations.
Also our construction is different from \cite{Kapustka}.
We construct the complete intersection compact
CY manifolds for all del Pezzo singularities.
This construction doesn't contradict the statement that for $n\leq 4$
the del Pezzo singularities are not complete intersections.
The price we have to pay is that these singularities will not be
generic, i.e. they will not have the maximal number of complex
deformations.
Whereas for the del Pezzo singularities with $n\geq 5$ and for
$\PP^1\times\PP^1$ we will represent all complex deformations.

\subsection{General construction}

At first we present the construction in the case of $dP_6$
singularity, and then give a more general formulation.
The input data is the embedding of $dP_6$ surface
in $\PP^3$ via a degree three equation.
The problem is to
find a CY threefold such that it has a local $dP_6$ singularity.
The solution has several steps
\begin{enumerate}
\item
Find the canonical class on $B=dP_6$ in terms of a restriction of a class
on $\PP^3$.
Let us denote this class as $K\in H^{1,1}(\PP^3)$.
$K$ can be found from expanding the total Chern class of $B$
\be
c(B)=\frac{(1+H)^4}{1+3H}=1+H+\ldots
\ee
thus $K=-c_1(B)=-H$.

\item
Construct the $\PP^1$ fiber bundle over $\PP^3$ as the
projectivisation
$M=P(\OO_{\PP^3}\oplus\OO_{\PP^3}(K))$.

\item
The Calabi-Yau $Y_3$ is given by an equation of degree $3$ in $\PP^3$ and
degree $2$ in 
the coordinates on the fiber.
The total Chern class of $Y_3$ is
\be
c(Y_3)=\frac{(1+H)^4(1+G-H)(1+G)}{1+3H+2G}
\ee
this has a vanishing first Chern class.
By construction, this Calabi-Yau has a del Pezzo singularity at
$t=0$.
\end{enumerate}

This construction has a generalization for the other del
Pezzo surfaces.
Let $B$ denote a del Pezzo surface
embedded in $X$ as a
complete intersection of a system of equations \cite{Hubsch}.
Assume, for concreteness, that the system contains two
equations and
denote by $L_1$ and $L_2$ the classes corresponding to the divisors
for these two equations in $X$.
The case of other number of equations can be obtained as a
straightforward generalization.

\begin{enumerate}
\item
First we find the canonical class of surface $B\subset X$,
defined in terms of two
equations with the corresponding classes $L_1,\:L_2\in H^{1,1}(X)$,
\be
c(B)=\frac{c(X)}{(1+L_1)(1+L_2)}=1+c_1(X)-L_1-L_2+\ldots
\ee
thus the canonical class of $X$ is obtained by the restriction of
$K=L_1+L_2-c_1(X)$.

\item
Second, we construct the $\PP^1$ fiber bundle over $X$ as the
projectivisation
$M=P(\OO_{X}\oplus\OO_{X}(K))$.

\item
In the case of two equations,
the Calabi-Yau manifold $Y_3\subset M$ is not unique.
Let $G$ be the hyperplane class in the fibers,
then we can write three different systems of equations that define a
CY manifold: the classes for the equations in the
first system are $L_1+2G$ and $L_2$, the second one
has $L_1+G$ and $L_2+G$, the third one has $L_1$ and $L_2+2G$.%
\footnote{
Here $L_1,L_2\in H^{1,1}(M)$ are defined via the pull back of the
corresponding classes in $H^{1,1}(X)$ with respect to the projection
of $\PP^1$ the fibers $\pi:M\ra X$.
}

As an example, let us describe the first system.
The first equation in this system is given by $L_1$ in $X$ and has degree $2$ in
the coordinates on the fibers.
The second equation is $L_2$ in $X$.
The total Chern class is
\be
c(Y_3)=\frac{c(X)(1+G+K)(1+G)}{(1+L_1+2G)(1+L_2)}.
\ee
Since $K=L_1+L_2-c_1(X)$, it is straightforward to check that the
first Chern class of $Y_3$ is trivial.
\end{enumerate}

Let us show how this program works in an example of a CY cone over
$B=\PP^1\times\PP^1$.
The $\PP^1\times\PP^1$ surface can be embedded in
$\PP^3$ by a generic degree two polynomial equation
\cite{Hubsch}\cite{Griffiths}
\be
P_2(z_i)=0
\ee
where $(z_0,\ldots,z_3)\in\PP^3$.
\footnote{
By coordinate
redefinition in $\PP^3$ one can represent the equation as
$z_0z_3=z_1z_2$.
The solutions of this equation can be parameterized by the points
$(x_1,y_1)\times(x_2,y_2)\in\PP^1\times\PP^1$ as
$
(z_0,z_1,z_2,z_3)=(x_1x_2, x_1y_2, y_1x_2, y_1y_2)
$.
This is the Segre embedding $\PP^1\times\PP^1\subset\PP^3$.
}

The first step of the program is to find the canonical class of $B$
in terms of a class in $\PP^3$.
Let $H$ be the hyperplane class of $\PP^3$.
Then the total Chern class of $B$ is
\be
c(B)=\frac{(1+H)^4}{1+2H}=1+2H+2H^2.
\ee
The canonical class is
\be
K(B)=-c_1(B)=-2H
\ee

Next we construct the $\PP^1$ bundle $M=P(\OO_{\PP^3}\oplus\OO_{\PP^3}(K))$ with the
coordinates $(s,t)$ along the fibers, where locally $s\in\OO_{\PP^3}$ and
$t\in\OO_{\PP^3}(-2H)$.
The equation that describes the embedding of the CY manifold $Y_3$
in $M$ is
\be
P_2(z_i)s^2+P_4(z_i)st+P_6(z_i)t^2=0
\ee
This equation is homogeneous in $z_i$ of degree two, since $t$ has
degree $-2$.
The section of $M$ at $t=0$ is contractable and the intersection
with the $Y_3$ is $P_2(z_i)=0$, i.e. $Y_3$ is the CY cone over
$\PP^1\times\PP^1$ near $t=0$.
The total Chern class of $Y_3$ is
\be
c(Y_3)=\frac{(1+H)^4(1+G)(1+G-2H)}{1+2H+2G}
\ee
It's easy to check that $c_1(Y_3)=0$.

\subsection{A discussion of deformations}

In this subsection we will discuss the deformations of the
del Pezzo singularities in the compact CY spaces.
The explicit description of the singularities and their deformations
can be found in the appendix.

The procedure is similar to the deformation of the $dP_6$
singularity described in section \ref{seqdelp}.
As before let $Y_3\subset M$ be an embedding of the CY threefold
$Y_3$ in $M$,
a $\PP^1$ bundle over products of (weighted) projective spaces.
If we blow down the section of the $\PP^1$ bundle that contains
the del Pezzo, then $M$ becomes a toric variety that we denote by
$V$.
After the blow down, equation for the CY in $M$ becomes a singular
equation for a CY embedded in $V$.
The last step is to deform the equation in $V$ to get a generic CY.%
\footnote{
In the example of $dP_6$ singularity on the quintic,
the projective bundle is
$M=P(\OO_{\PP^3}\oplus\OO_{\PP^3}(-H))$,
the manifold $V$\!, obtained by blowing down the exceptional $\PP^3$ in
$M$\!, is $\PP^4$\!, and the singular equation is the
singular quintic in $\PP^4$.
}

Let $n$ denote the
number of two-cycles on del Pezzo with self intersection $(-2)$.
The intersection matrix of these
cycles is minus the Cartan matrix of the corresponding Lie algebra $E_n$.
The maximal number of complex deformations of del Pezzo singularity is
$c^\vee(E_n)-1$, where $c^\vee(E_n)$ is the dual Coxeter number of
$E_n$.
These deformations can be performed only if the del Pezzo has a
zero size.
As a result of these deformations the canonical class on the del
Pezzo becomes trivial within the CY and the del Pezzo singularity
is partially or completely smoothed out.
In the generic situation we expect that all $(-2)$ two-cycles on del
Pezzo are trivial within the CY, then the number of complex
deformations is maximal (this will be the case for
$\PP^1\times\PP^1$, $dP_5$, $dP_6$, $dP_7$, $dP_8$).
If some of the $(-2)$ two-cycles become non trivial within the CY,
then the number of complex deformations of the corresponding cone is
smaller.
We will observe this for our embedding of $dP_2$, $dP_3$,
and $dP_4$.
This reduction of the number of complex deformations depends on the
particular embedding of del Pezzo cone.
In \cite{Aganagic:2006ex}, the generic deformations of the cones over
$dP_2$ and $dP_3$ were constructed.
The list of embeddings of del Pezzo singularities and their
deformations can be found in the Appendix.
The results on the number of complex deformations and the comparison
with the maximal number of deformations $(c^\vee-1)$ are presented in the tables
below.

{\it Table 1. Some characteristics of del Pezzo surfaces.}

\nopagebreak
{\small\medskip
\noindent
\begin{tabular}{|c|c|c|c|c|c|c|c|}
  \hline
  del Pezzo & \# two-cycles & \# (-2) two-cycles & Dynkin diagram & $c^\vee-1$  \\
  \hline
  $\PP^2$ & 1 & 0 & 0 & 0 \\
  $\PP^1\times\PP^1$ & 2 & 1 & $A_1$ & 1 \\
  $dP_1$ & 2 & 0 & 0 & 0 \\
  $dP_2$ & 3 & 1 & $A_1$ & 1 \\
  $dP_3$ & 4 & 3 & $A_2\times A_1$ & 3 \\
  $dP_4$ & 5 & 4 & $A_4$ & 4 \\
  $dP_5$ & 6 & 5 & $D_5$ & 7 \\
  $dP_6$ & 7 & 6 & $E_6$ & 11 \\
  $dP_7$ & 8 & 7 & $E_7$ & 17 \\
  $dP_8$ & 9 & 8 & $E_8$ & 29 \\
  \hline
\end{tabular}
}

\bigskip

{\it Table 2. 
Complex deformations of del Pezzo singularities studied in the
paper}
\nopagebreak

{\small\medskip
\noindent
\begin{tabular}{|c|c|c|c|c|c|}
  \hline
  del Pezzo & \# (-2) two-cycles & \# trivial (-2) two-cycles & $c^\vee-1$ & \# complex deforms \\
  \hline
  $\PP^2$ & 0 & 0 & 0 & 0 \\
  $\PP^1\times\PP^1$ & 1 & 1 & 1 &  1 \\
  $dP_1$ & 0 & 0 & 0 & 0 \\
  $dP_2$ & 1 & 0 & 1 & 0 \\
  $dP_3$ & 3 & 1 & 3 & 1 \\
  $dP_4$ & 4 & 3 & 4 & 3 \\
  $dP_5$ & 5 & 5 & 7 & 7 \\
  $dP_6$ & 6 & 6 & 11 & 11 \\
  $dP_7$ & 7 & 7 & 17 & 17 \\
  $dP_8$ & 8 & 8 & 29 & 29 \\
  \hline
\end{tabular}
}
\bigskip

\section{Conclusions and outlook}

In this paper we have constructed a class of compact Calabi-Yau manifolds
that have del Pezzo singularities.
The construction is analytic, i.e. the CY manifolds are described by
a system of equations in the $\PP^1$ bundles over the projective
spaces.

We argue that this construction can be used for the geometrical SUSY
breaking \cite{Aganagic:2006ex} as well as for the compactification
of ISS \cite{ISS}.
As an example, we find a compact CY manifold with del Pezzo $6$
singularity and some conifolds such that some 2-cycles on del Pezzo
are homologous to the 2-cycles on the conifolds.
Also we find an ISS vacuum in the quiver gauge theory for $dP_6$
singularity.

In the last section and in the Appendix,
we describe the deformations of del Pezzo singularities.
The del Pezzo $n$ surface corresponds to the Lie group $E_n$.
The expected number of complex deformations for the cone over del
Pezzo is $c^\vee(E_n)-1$, where $c^\vee$ is the dual Coxeter number for
the Lie group $E_n$.
In the studied examples, the cones over $\PP^1\times\PP^1$ and
over $dP_5$, $dP_6$, $dP_7$, and $dP_8$ have generic deformations.
But the cones over $dP_2$, $dP_3$ and $dP_4$ have less deformations,
i.e. these cones do not describe the most generic embedding of the
corresponding del Pezzo singularities.%
\footnote{
It is known that the generic embeddings of del Pezzo $n$ singularities
for $n\leq 4$ (or rank $k=9-n\geq 5$)
cannot be represented as complete intersections
\cite{Gross}\cite{Reid-dP}, in our construction the del Pezzo
singularities are non generic complete intersections.
}

We propose that for the generic embedding the two-cycles on del
Pezzo with self-intersection $(-2)$ are trivial within the full
Calabi-Yau geometry.
The non trivial two cycles with self-intersection $(-2)$ impose
restrictions on the complex deformations.
This proposal agrees with the above examples of the embeddings
of del Pezzo singularities.
Also we get a similar conclusion when the CY has some number of
conifolds in addition to the del Pezzo singularity.
Although the conifolds are away from the del Pezzo and the del
Pezzo itself is not singular, it acquires a non trivial two-cycle
and the number of deformations is reduced.

Sometimes the F-theory/orientifolds point of view has advantages
compared to the type IIB theory.
Our construction of CY threefolds can be generalized to find the
3-dimensional base spaces of elliptic fibrations in F-theory with
the necessary del Pezzo singularities.
Also we expect this construction to be useful as a first step in finding
the warped deformations of the del Pezzo singularities and in
the studies of the Landscape of string compactifications.

\bigskip
\bigskip

{\bf Acknowledgements.}

The author is thankful to Herman Verlinde for the close attention
to the work, to Igor Klebanov,
Nikita Nekrasov, Matt Buican, Sebastian Franco, Yanir Rubinstein,
Sergio Benvenuti, and
Yuji Tachikawa for valuable discussions and comments.
The work is supported in part by Russian Foundation of Basic
Research under the grant RFBR 06-02-17383.

\bigskip
\bigskip
\bigskip
\bigskip
\bigskip

\newpage

{\bf
Appendix. A list of compact CY with del Pezzo singularities}
\bigskip

In the appendix we construct the embeddings of all del Pezzo
singularities in compact CY manifolds and describe the complex
deformations of these embeddings.
This description follows the general construction in section 4.

In the following $B$ denotes the two-dimensional del Pezzo surface and
$X$ denotes the space where we embed $B$.
The space $X$ will be either a product of projective spaces or a weighted
projective space.
For example, if $B\subset X=\PP^n\times\PP^m\times\PP^k$, then the
coordinates on the three projective spaces will be denoted as
$(z_0,\ldots,z_n)$, $(u_0,\ldots,u_m)$, and $(v_0,\ldots,v_k)$
respectively.
The hyperplane classes of the three projective spaces will be
denoted by $H$, $K$, $R$ respectively.

A polynomial of degree $q$ in $z_i$, degree $r$ in $u_j$, and degree
$s$ in $v_l$ will be denoted by $P_{q,r,s}(z_i;u_j;v_l)$.

If there are only two or one projective space, then we will use the
first two or the first one projective spaces in the above
definitions.

For the weighted projective spaces, we will use the notations of
\cite{Friedman:1997yq}.
For example, consider the space
$W\PP^3_{11pq}$, where $p,q\in{\NN}$.
The dimension of this space is $3$,
the subscripts $(1,1,p,q)$ denote the weights
of the coordinates with respect to
the projective identifications
$(z_0,z_1,z_2,z_3)\sim (\ld z_0,\ld z_1,\ld^p z_2,\ld^q z_3)$.

The $\PP^1$ bundles over $X$ will be denoted as
$M=P(\OO_X\oplus\OO_X(K))$,
where $K$ is the class on $X$ that restricts to the
canonical class on $B$.
The coordinates on the fibers will be $(s,t)$ so that locally
$s\in\OO_X$ and $t\in\OO_X(K)$.
The hyperplane class of the fibers will be denoted by $G$, it
satisfies the property $G(G+K)=0$ for $M=P(\OO_X\oplus\OO_X(K))$.
In the construction of the $\PP^1$ bundles,
we will use the fact that $K(B)=-c_1(B)$ and
will not calculate $K(B)$ separately.


The deformations of some del Pezzo singularities will be described via
embedding in particular toric varieties.
We will call them generalized weighted projective spaces.
Consider, for example, the following notation
\be
GW\PP^5{}^{^{}_{11100002}}_{^{00011001}_{00000111}}
\ee
The number $5$ is the dimension of the space. This space is obtained
from ${\C^8}^*$ by taking the classes of equivalence with respect to
three identifications.
The numbers in the three rows correspond to the charges under these
identifications.
\bea
(z_1,z_2,z_3,z_4,z_5,z_6,z_7,z_8)\sim
(\ld_1z_1,\ld_1z_2,\ld_1z_3,z_4,z_5,z_6,z_7,\ld_1^2z_8)\\
(z_1,z_2,z_3,z_4,z_5,z_6,z_7,z_8)\sim
(z_1,z_2,z_3,\ld_2z_4,\ld_2z_5,z_6,z_7,\ld_2z_8)\\
(z_1,z_2,z_3,z_4,z_5,z_6,z_7,z_8)\sim
(z_1,z_2,z_3,z_4,z_5,\ld_3z_6,\ld_3z_7,\ld_3z_8)
\eea


\begin{enumerate}
    \item $B=\PP^2\subset X=\PP^3$.\\
The equation for $B$
\be
P_1(z_i)=0.
\ee
The total Chern class of $B$
\be
c(B)=(1+H)^3
=1+3H+3H^2
\ee
The $\PP^1$ bundle is $M=P(\OO_X\oplus\OO_X(-3H))$.
The equation for the Calabi-Yau threefold $Y_3$
\be
P_1(z_i)s^2+P_4(z_i)st+P_7(z_i)t^2=0.
\ee
The total Chern class of $Y_3$ is
\be
c(Y_3)=\frac{(1+H)^4(1+G)(1+G-3H)}{1+H+2G}.
\ee
It is easy to see that the first Chern class is zero.
The calculation of the Chern classes for the CY manifolds that we
present below is similar and we will not repeat it.

The embedding space $V=W\PP^4_{11113}$ has the coordinates $(z_0,\ldots,z_3;w)$
and the singular CY is
\be
P_1(z_0,\ldots,z_3)w^2+P_4(z_0,\ldots,z_3)w+P_7(z_0,\ldots,z_3)=0
\ee
This is already the most general equation, i.e. there are no additional
complex deformations.

    \item $B=\PP^1\times\PP^1\subset X=\PP^3$.\\
The equation for $B$
\be
P_2(z_i)=0.
\ee
The total Chern class of $B$
\be
c(B)=\frac{(1+H)^4}{1+2H}
=1+2H+2H^2
\ee
The $\PP^1$ bundle is $M=P(\OO_X\oplus\OO_X(-2H))$.
The equation for the Calabi-Yau threefold $Y_3$
\be
P_2(z_i)s^2+P_4(z_i)st+P_6(z_i)t^2=0
\ee
The embedding space $V=W\PP^4_{11112}$ has the coordinates $(z_0,\ldots,z_3;w)$
and the singular CY is
\be
P_2(z_i)w^2+P_4(z_i)w+P_6(z_i)=0
\ee
This equation has one deformation $k w^3$ and the spaces $M$ and
$V$ have the same number of coordinate redefinitions.
Thus the space of complex deformations is one-dimensional.

    \item $B=dP_1\subset X=\PP^2\times\PP^1$\\
The equation defining $B$ has degree one in $z_i$ and degree one in
$u_j$
\be
P_1(z_i)u_0+Q_1(z_i)u_1=0.
\ee
The total Chern class of $B$
\be
c(B)=\frac{(1+H)^3(1+K)^2}{1+H+K}
=1+2H+K+H^2+3HK
\ee
The $\PP^1$ bundle is $M=P(\OO_X\oplus\OO_X(-2H-K))$.
The equation for the Calabi-Yau threefold $Y_3$ is
\be
P_{1,1}(z_i;u_j)s^2+P_{3,2}(z_i;u_j)st+P_{5,3}(z_i;u_j)t^2=0
\ee
The embedding space $V=GW\PP^4_{^{111002}_{000111}}$ has the coordinates
$(z_0,z_1,z_2;u_0,u_1;w)$
and the singular CY is
\be
P_{1,1}(z_i;u_j)w^2+P_{3,2}(z_i;u_j)w+P_{5,3}(z_i;u_j)=0
\ee
There are no complex deformations of this equation.

    \item $B=dP_2\subset X=\PP^2\times\PP^1\times\PP^1$\\
The del Pezzo surface is defined by a system of two equations.
The first equation has degree one in $z_i$ and degree one in $u_k$.
The second equation has degree one in $z_i$ and degree one in $v_k$.
\bea
P_1(z_i)u_0+Q_1(z_i)u_1=0\\
R_1(z_i)v_0+S_1(z_i)v_1=0
\eea
The total Chern class of $B$
\be
c(B)=\frac{(1+H)^3(1+K)^2(1+R)^2}{(1+H+K)(1+H+R)}
=1+2H+K+R+2H(K+R)+KR\\
\ee
The $\PP^1$ bundle is $M=P(\OO_X\oplus\OO_X(-2H-K-R))$.
The system of equations for the Calabi-Yau threefold $Y_3$ can be
written as
\bea
P_{1,1,0}(z_i;u_k;v_k)s^2+P_{3,2,1}(z_i;u_k;v_k)st+P_{5,3,2}(z_i;u_k;v_k)t^2=0\\
Q_{1,0,1}(z_i;u_k;v_k)=0
\eea
The space $V=GW\PP^5{}^{^{}_{11100002}}_{^{00011001}_{00000111}}$
has the coordinates $(z_0,z_1,z_2;u_0,u_1;v_0,v_1;w)$
and the singular CY is
\bea
P_{1,1,0}(z_i;u_k;v_k)w^2+P_{3,2,1}(z_i;u_k;v_k)w+P_{5,3,2}(z_i;u_k;v_k)=0\\
Q_{1,0,1}(z_i;u_k;v_k)=0
\eea
There are no complex deformations of this equation.
This is in contradiction with the general expectation of one complex
deformation, i.e. the embedding is not the most general.
This is connected to the fact that all the two-cycles on the del
Pezzo are non trivial within the CY.

    \item $B=dP_3\subset X=\PP^1\times\PP^1\times\PP^1$\\
The del Pezzo surface is defined by an equation
of degree one in $z_i$, degree one in $u_j$ and degree one in $v_k$.
\be
P_{1,1,1}(z_i;u_j;v_k)=0
\ee
The total Chern class of $B$
\be
c(B)=\frac{(1+H)^2(1+K)^2(1+R)^2}{(1+H+K+R)}
=1+(H+K+R)+2(HK+HR+KR)\\
\ee
where $H$, $K$ and $R$ are the hyperplane classes on
the three $\PP^1$'s.
The $\PP^1$ bundle is $M=P(\OO_X\oplus\OO_X(-H-K-R))$.
The equation for the Calabi-Yau threefold $Y_3$ is
\be
P_{1,1,1}(z_i;u_j;v_k)s^2+P_{2,2,2}(z_i;u_j;v_k)st+P_{3,3,3}(z_i;u_j;v_k)t^2=0
\ee
The embedding space $V=GW\PP^4{}^{^{}_{1100001}}_{^{0011001}_{0000111}}$
has the coordinates $(z_0,z_1;u_0,u_1;v_0,v_1;w)$
and the singular CY is
\be
P_{1,1,1}(z_i;u_j;v_k)w^2+P_{2,2,2}(z_i;u_j;v_k)w+P_{3,3,3}(z_i;u_j;v_k)=0
\ee
This equation has one deformation $kw^3$ and the spaces $M$ and $V$
have the same number of reparameterizations.
Consequently, there is
one complex deformation of the cone.
This is related to the fact that 3 out of 4 two-cycles on $dP_3$ are
independent within the CY and there is only one $(-2)$ two-cycle on
$dP_3$ that is trivial within the CY.


    \item $B=dP_4\subset X=\PP^2\times\PP^1$\\
Equation defining $B$ has degree two in $z_i$ and degree one in
$u_j$
\be
P_2(z_i)u_0+Q_2(z_i)u_1=0.
\ee
The total Chern class of $B$
\be
c(B)=\frac{(1+H)^3(1+K)^2}{1+2H+K}
=1+H+K+H^2+3HK\\
\ee
where $H$ and $K$ are the hyperplane classes on $\PP^2$ and $\PP^1$
respectively.
The $\PP^1$ bundle is $M=P(\OO_X\oplus\OO_X(-H-K))$.
The equation for the Calabi-Yau threefold $Y_3$ is
\be
P_{2,1}(z_i;u_j)s^2+P_{3,2}(z_i;u_j)st+P_{4,3}(z_i;u_j)t^2=0
\ee
The embedding space $V=GW\PP^4_{^{111001}_{000111}}$
has the coordinates $(z_0,z_1,z_3;u_0,u_1;w)$
and the singular CY is
\be
P_{2,1}(z_i;u_j)w^2+P_{3,2}(z_i;u_j)w+P_{4,3}(z_i;u_j)=0
\ee
The deformations of the singularity have the form of degree one
polynomial in $z_0,z_1,z_2$ times $w^3$.
Consequently, there are three deformation
parameters and the spaces $V$ and $M$ have the same
reparameterizations.
In this case we have three complex deformations and three $(-2)$
two-cycles on $dP_4$ that are trivial within CY.

    \item $B=dP_5\subset X=\PP^4$.\\
The del Pezzo surface is defined by a system of two equations.
Both equation have degree $2$ in $z_i$.
\bea
P_2(z_i)=0\\
R_2(z_i)=0
\eea
The total Chern class of $B$
\be
c(B)=\frac{(1+H)^5}{(1+2H)^2}
=1+H+2H^2\\
\ee
The $\PP^1$ bundle is $M=P(\OO_X\oplus\OO_X(-H))$.
The system of equations for the first possible
Calabi-Yau threefold $Y_3$ is
\bea
P_2(z_i)s^2+P_3(z_i)st+P_4(z_i)t^2=0\\
R_2(z_i)=0
\eea
It has the following characteristics
\bea
\chi(Y_3)=-160;\\
h^{1,1}(Y_3)=2;\\
h^{2,1}=82.
\eea

Now we find the deformations of this cone over $dP_5$.
The $\PP^1$ bundle $M$ is, in fact, the $\PP^5$ blown up at one
point.
By blowing down the $t=0$ section of $M$ we get $\PP^5$.
The $CY$ three-fold with the $dP_5$ singularity
is embedded in $\PP^5$ by the system of two
equations
\be\lb{conedP5}
\ba{r}
P_2(z_i)w^2+P_3(z_i)w+P_4(z_i)=0;\\
R_2(z_i)=0.
\ea
\ee
The deformations of the singularity correspond to taking a general
degree four polynomial in the first equation.
This general CY has
\bea
\chi=-176;\\
h^{1,1}(Y_3)=1;\\
h^{2,1}=89.
\eea
Since the system (\ref{conedP5}) has only the $dP_5$ singularity and
the general CY manifold has $89-82=7$ more complex deformations, we
interpret these extra $7$ deformations as the deformations of the
cone over $dP_5$.
This number is consistent with the general expectation, since
$c^\vee(D5)-1=7$, where $c^\vee(D5)=8$ is the dual Coxeter number
for $D5$.

The second CY with the $dP_5$ singularity is described by
\bea
P_2(z_i)s+P_3(z_i)t=0\\
R_2(z_i)s+R_3(z_i)t=0.
\eea
Using the same methods as for the first CY, one can show that this
singularity also has $7$ complex deformations.

    \item $B=dP_6\subset X=\PP^3$.\\
The case of $dP_6$ was described in details section 2,
here we just repeat the general results.

The equation defining $dP_6\subset\PP^3$
\be
P_3(z_i)=0.
\ee
The total Chern class of $B$
\be
c(B)=\frac{(1+H)^4}{1+3H}
=1+H+3H^2
\ee
The $\PP^1$ bundle is $M=P(\OO_X\oplus\OO_X(-H))$.

The equation for the Calabi-Yau threefold $Y_3$
\be
P_3(z_i)s^2+P_4(z_i)st+P_5(z_i)t^2=0
\ee
The Euler number and the cohomologies of $Y_3$ are
\bea
\chi=-176\\
h^{1,1}=2\\
h^{2,1}=90
\eea
The deformation of this singularity is a quintic in $\PP^4$, that
has
\be
h^{2,1}=101
\ee
complex deformations.
The difference between the number of complex deformations is
$101-90=11$, which is consistent with $c^\vee(E6)-1=11$.

    \item $B=dP_7\subset X= W\PP^3_{1112}$.\\
The equation defining $B$ is homogeneous of degree four in $z_i$'s
\be
P_4(z_i)=0.
\ee
The total Chern class of $B$
\be
c(B)=\frac{(1+H)^3(1+2H)}{1+4H}
=1+H+5H^2
\ee
The $\PP^1$ bundle is $M=P(\OO_X\oplus\OO_X(-H))$.
The equation for the Calabi-Yau threefold $Y_3$
\be
P_4(z_i)s^2+P_5(z_i)st+P_6(z_i)t^2=0
\ee
The Euler number and the cohomologies of $Y_3$ are
\bea
\chi=-168\\
h^{1,1}=2\\
h^{2,1}=86
\eea
Blowing down the $t=0$ section of $M$ we get $V=W\PP^4_{11112}$.
The general CY is given by the degree six equation in $V$.
The total Chern class of this CY is
\be
c=\frac{(1+H)^4(1+2H)}{(1+6H)}
\ee
And the number of complex deformations
\be
h^{2,1}=103
\ee
The difference $103-86=17$ is equal to $c^\vee(E7)-1=17$,
where $c^\vee(E7)=18$ is the dual Coxeter number of $E7$.
Consequently, we can represent all complex deformations of $dP_7$
singularity in this embedding.

    \item $B=dP_8\subset X= W\PP^3_{1123}$.\\
The equation defining $B$ has degree six
\be
P_6(z_i)=0.
\ee
The total Chern class of $B$
\be
c(B)=\frac{(1+H)^2(1+2H)(1+3H)}{1+6H}
=1+H+11H^2
\ee
The $\PP^1$ bundle is $M=P(\OO_X\oplus\OO_X(-H))$.
The equation for the Calabi-Yau threefold $Y_3$
\be
P_6(z_i)s^2+P_7(z_i)st+P_8(z_i)t^2=0
\ee
The problem with this CY is that for any polynomials $P_6$, $P_7$
and $P_8$ it has a singularity at $s=z_0=z_1=z_2=z_3=0$ and $z_4=1$.
As a consequence the naive calculation of the Euler number gives a
fractional number
\be
\chi=-150\frac{2}{3}.
\ee
The good feature of this singularity is that it is away from the
del Pezzo, thus one can argue that this singularity should not affect
the deformation of the $dP_8$ cone.
In order to justify that we calculate the number of complex
deformations of the CY manifold with $dP_8$ singulariy
by calculating
the number of coefficients in the equation
minus the number of reparamterizations of
$M$.
The result is
\be
h^{2,1}=77.
\ee

Blowing down the $t=0$ section of $M$ we get $V=W\PP^4_{11123}$.
The general CY is given by the degree eight equation in $V$.
The number of coefficients
minus the number of reparamterizations of
$V=W\PP^4_{11123}$ is
\be
h^{2,1}=106.
\ee
The difference $106-77=29$ is equal to $c^\vee(E8)-1=29$,
where $c^\vee(E8)=30$ is the dual Coxeter number of $E8$.
Thus all complex deformations of $dP_8$ singularity
can be realized in this embedding.

\end{enumerate}

\end{document}